\documentclass[12pt]{iopart}
\usepackage{amssymb}
\usepackage{epsfig}
\usepackage[caption=false]{subfig}

\begin{document}

\title{Wigner transport equation with finite coherence length}

\author{Carlo Jacoboni, Paolo Bordone} 
%\author{Paolo Bordone}

\address{Dipartimento di Scienze Fisiche, Informatiche e Matematiche,
  Universit\`{a} di Modena e Reggio Emilia, and Centro S3, CNR-Istituto di Nanoscienze,Via Campi 213/A,  Modena I-41125,
  Italy}
\ead{bordone@unimore.it}
\begin{abstract}
The use of the Wigner function for the study of quantum transport in open systems is subject to severe criticisms. 
Some of the problems arise from the assumption of infinite coherence length of the electron dynamics outside the system of interest. 
In the present work the theory of the Wigner function is revised assuming a finite coherence length. A new dynamical equation
is found, corresponding to move the Wigner momentum off the real axis, and a numerical analysis is performed for the case 
of study of the one-dimensional potential barrier. In quantum device simulations, for a sufficiently long coherence length,
the new formulation does not modify the physics in any finite region of interest but it prevents mathematical divergence problems.
\end{abstract}

\maketitle

\section{Introduction}
Nowadays sophisticated technologies produce physical systems, and in particular semiconductor devices, of 
dimensions comparable with the electron wavelength or with the electron coherence length. Under such conditions,
semiclassical dynamics is not justified in principle, and a full quantum analysis has to be considered.
Among the possible different approaches, the Wigner function (WF) has been widely employed. This function has been
succesfully used in several fields of quantum statistical physics, such as molecular, atomic, and nuclear physics, quantum optics, 
quantum chemistry, quantum entanglement and 
entropy~\cite{carr83,monteiro94,manfredi00,fan03,cancellieri07,buscemi08,jacoboni10}\footnote{The Journal of Optics B
has published a Wigner Centennial issue (J. Optics B {\bf 5}(3), (2003)), where many references can be found}.
In particular, the WF has proved to be very useful for studying  quantum electron transport~\cite{frensley86,kriman87,kluksdahl89,frensley90,bout90,ferry93,nedjalkov97,pascoli98,bordone99,bertoni99,barletti01,jacoboni01,jacoboni04,nedjalkov04,demeio05,nedjalkov06,querlioz08,morandi09,trovato11},
owing to its strong analogy with the semiclassical picture, since it explicitly refers to variables defined in an ({\bf r},{\bf p}) 
Wigner phase space, together with a rigorous description of electron dynamics in quantum terms. The WF has been 
considered especially suitable for studying mesoscopic systems. Their typical dimensions are such that transport cannot be assumed to be totally
coherent since dissipative scattering begins to take place. In such a condition, the Schr\"{o}dinger equation for isolated electrons 
cannot be used. On the other hand, dimensions are so small that coherent quantum effects are present, and the system 
does not present within itself a sufficiently large number of microscopic situations to justify configuration averages.
Furthermore, the WF has been considered to be an appropriate tool to treat the problem of the contacts in electronic devices. In fact, being defined in a 
phase space, both position and momentum can be considered simultaneously, and the connection with semiclassical systems 
described by a classical distribution in the contacts should be straightforward. Usually, the perturbation theory is used in dealing with
quantum electron transport. The form of the quantum equation describing the dynamics of the electron system depends on the
choice of the unperturbed Hamiltonian $H_0$. For example $H_0$ can describe the free electron dynamics, or it can also include
the potential profile, or part of it. The corresponding set of eigenstates of $H_0$ is used to obtain the appropriate form
of the evolution equation and to define the initial boundary conditions. An electronic device with contacts is an open system, 
and it has been recognized that a natural basis of quantum states for such systems is provided by the so-called 
``scattering states''~\cite{kriman87,nedjalkov97}.

Open systems, however, pose several problems to the use of the WF for studying quantum transport. Basis eigenstates for 
such systems extend to infinity, and the very definition of the WF must pay the price of dealing with improper 
functions~\cite{nedjalkov97}. Indeed, in itself, such a problem could not be dramatic: the use of improper functions in
quantum mechanics is well established when dealing with extended states. However, a second problem arises in 
the derivation of the equation for the WF. During such a derivation an integration by parts is needed~\cite{jacoboni04} where 
one term is canceled if the wave function vanishes to infinity. If this is not the case, as in the coherent limit (extended states in absence
of collisions), neglecting this term is not justified~\footnote{An alternative derivation seems to avoid this problem (private communication
by F. Rossi and by L. Demeio), but in this case we need to assume the existence of the Fourier transform and
of its anti-transform in a generalised sense, and this is not always rigorously justified.}. 
\par
Furthermore, in recent times, Fausto Rossi and coworkers have published a number of papers~\cite{zaccaria03,taj06,rosati13} where
they seriously question the use of the WF for quantum transport in open systems. In particular, in~\cite{rosati13}, it is pointed out that
i) in the coherent limit, with inflow boundary conditions, the solution may be not 
unique~\footnote{This result must be compared with a previous paper~\cite{barletti01}, where different conclusions
are reached.},
and ii) the exact boundary conditions must be known in order to get a correct physical solution inside the device of interest.
However, such precise knowledge is illusory, since the solution of the problem should already be known, 
and the use of approximate (e.g. semiclassical) boundary conditions can lead to unphysical solutions, namely the particle density 
can become negative.\par
Most of the above difficulties originate from the infinite coherence length of the basis extended states, as e.g. the scattering states.
To shed new light  on this issue it is useful to analyze the problem from scratch by assuming a given finite coherence 
length $\lambda$ in the definition of the WF~\footnote{For numerical purposes a finite coherence length has been sometimes 
introduced in the potential term of the WF, see e.g.~\cite{jiang11,spisak13}.}. 
In real systems scattering processes will take care of $\lambda$.
Pursuing this project, in this paper a new equation is derived, where the role of the coherence length is shown explicitly.
It is worth noting that introducing a finite coherence length corresponds to introduce an imaginary part of the
Wigner momentum, that means, in mathematical terms, to move it off the real axis,  in analogy with what is done with the energy in the 
Green function theory.

The general theory of the WF equation with finite coherence length, including its dynamical equation, is given in Sec. II. Sec. III 
discusses a number of properties of the new function, the limit of the dynamic equation when the coherence length tends to
infinity, end a numerical example for the case of study of the one-dimensional potential barrier. Conclusions are drawn in Sec. IV.

\section{Theory}
\label{sec:1}
\subsection{Finite coherence length}
Let us consider for simplicity a single-particle one-dimensional system. The usual definition of the WF is~\cite{jacoboni10}:
\begin{equation}
f_w(x,p,t)=\int_{-\infty}^{\infty}\ e^{-ips/\hbar}\ \overline{\Psi(x+s/2,t)\Psi^*(x-s/2,t)}\ ds\ ,
\end{equation}
where $\Psi(x,t)$ is the particle wave function, $p$ is the Wigner momentum, and the overbar indicates an ensemble 
average. The contribution to the above integral is significantly different from zero as long as the wavefunctions in the ensemble
maintains some coherence between two points at distance $s$. If a stationary system is considered, and we limit ourself to a 
single extended eigenstate $\psi(x)$, the 
coherence length extends to infinity and the WF becomes an improper function. In a real physical system the scattering processes
induce decoherence, thus  reducing the coherence length to a finite value. Here we mimic such effect by
introducing a coherence length $\lambda>0$ directly in the definition of the WF, that now reads:
\begin{equation}\label{WFc}
f_{wc}(x,p)=\int_{-\infty}^{\infty}\ e^{-ips/\hbar}\ e^{-\vert s\vert/\lambda}\ \psi(x+s/2)\psi^*(x-s/2)\ ds\ .
\end{equation}
%Following the usual procedure it is possible to obtain the dynamical equation:
%\begin{eqnarray}\label{WEc}
% i\hbar\frac{\partial f_{wc}}{\partial t}&=&\int_{-\infty}^{\infty}\ e^{-ips/\hbar}\ e^{-\vert s\vert/\lambda}\ 
%\left[\left(i\hbar\frac{\partial}{\partial t}\psi(+)\right)\psi^*(-)+\psi(+)\left(i\hbar\frac{\partial}{\partial t}\psi^*(-)\right)\right]\ ds\nonumber \\ 
%&=&\int_{-\infty}^{\infty}\ e^{-ips/\hbar}\ e^{-\vert s\vert/\lambda}\left[({\cal H}\psi(+))\psi^*(-)-\psi(+)({\cal H}\psi(-))^*\right]\ ds\ ,
%\end{eqnarray}
%where, for simplicity of notation, hereafter $(+)$ and $(-)$ stay for $(x+s/2)$ and $(x-s/2)$ respectively, and ${\cal H}$ is the system Hamiltonian
%\begin{equation}
%{\cal H}=-\frac{\hbar^2}{2m}\frac{\partial^2}{\partial x^2}+V(x)\ ;
%\end{equation}
%$m$ is the effective mass of the particle.
Let us consider the Hamiltonian ${\cal H}$ of the system 
\begin{equation}\label{ham}
{\cal H}=-\frac{\hbar^2}{2m}\frac{\partial^2}{\partial x^2}+V(x)\ ,
\end{equation}
where $m$ is the effective mass of the particle and $V(x)$ is the potential energy. Following the standard procedure we obtain 
the dynamical equation of the WF~\cite{jacoboni10}.
The kinetic term reads:
\begin{eqnarray}
&&\frac{\partial f_{wc}}{\partial t}\Bigg\vert_{kin}=\frac{2i\hbar}{m}\left\{\int_{-\infty}^0 e^{-ips/\hbar}\ e^{s/\lambda}
\frac{\partial}{\partial s}\left[\frac{\partial\psi(+)}{\partial s}\psi^*(-)-
\psi(+)\frac{\partial\psi^*(-)}{\partial s}\right]\ ds\right.\nonumber \\ 
&&\hspace{40 pt} \left.+\int_0^{\infty} e^{-ips/\hbar}\ e^{-s/\lambda}
\frac{\partial}{\partial s}\left[\frac{\partial\psi(+)}{\partial s}\psi^*(-)-\psi(+)\frac{\partial\psi^*(-)}{\partial s}\right]\ ds\right\}\ .
\end{eqnarray}
Integrating by parts the first terms of the integration vanish because of the damping factor (contrary to what happens
in the standard theory). The second terms of the integration by parts yield
\begin{eqnarray}\label{expr1}
&&\hspace{-30 pt}\frac{2i\hbar}{m}\left\{-\left(-\frac{ip}{\hbar}+\frac{1}{\lambda}\right)\int_{-\infty}^0 e^{-ips/\hbar}\ e^{s/\lambda}
\left[\frac{\partial\psi(+)}{\partial s}\psi^*(-)-\psi(+)\frac{\partial\psi^*(-)}{\partial s}\right]\right.\ ds\nonumber\\ 
&&\hspace{-30 pt}\left.-\left(-\frac{ip}{\hbar}-\frac{1}{\lambda}\right)\int_0^{\infty} e^{-ips/\hbar}\ e^{-s/\lambda}
\left[\frac{\partial\psi(+)}{\partial s}\psi^*(-)-\psi(+)\frac{\partial\psi^*(-)}{\partial s}\right]\right\}\ ds\ .
\end{eqnarray}
The first terms of each line in (\ref{expr1}) combine to give
\begin{equation}\label{primit}
 I_k^{(1)}=-\frac{2p}{m} 
\int_{-\infty}^\infty e^{-ips/\hbar}\ e^{-\vert s\vert/\lambda}\left[\frac{\partial\psi(+)}{\partial s}\psi^*(-)-\psi(+)\frac{\partial\psi^*(-)}{\partial s}\right]\ ds\ .
\end{equation}
Following again the usual procedure, from (\ref{primit}) we get a first contribution to the kinetic term
\begin{equation}\label{kin1}
 \frac{\partial f_{wc}}{\partial t}\Bigg\vert_{kin}^{(1)}=-\frac{p}{m}\frac{\partial}{\partial x}
\int_{-\infty}^\infty e^{-ips/\hbar}\ e^{-\vert s\vert/\lambda}\ \psi(+)\psi^*(-)\ ds 
=-\frac{p}{m}\frac{\partial f_{wc}}{\partial x}\ ,
\end{equation}
where the definition in equation (\ref{WFc}) has been used. This is the usual free-particle term of the Wigner transport equation.
Now, however, the second terms of each line in (\ref{expr1}) have to be considered. Straightforward calculations lead to:
\begin{eqnarray}
 \frac{\partial f_{wc}}{\partial t}\Bigg\vert_{kin}^{(2)}&=&
-\frac{i\hbar}{m\lambda}\frac{\partial}{\partial x}\left\{\int_{-\infty}^0 e^{-ips/\hbar}\ e^{s/\lambda}\ \psi(+)\psi^*(-)\ ds\right. \nonumber \\
&&\hspace{45 pt}-\left.\int_0^\infty e^{-ips/\hbar}\ e^{-s/\lambda}\ \psi(+)\psi^*(-)\ ds\right\} \ .
\end{eqnarray}
Substituting $s$ with $-s$ in the first integral of the above expression, a more compact form is achieved
\begin{equation}\label{kin2}
i\hbar\frac{\partial f_{wc}}{\partial t}\Bigg\vert_{kin}^{(2)}=2i\frac{\hbar^2}{m}\frac{1}{\lambda}\frac{\partial}{\partial x}\Im\int_0^{\infty}\ e^{-s/\lambda}e^{ips/\hbar}\ \psi(-)\psi^*(+)\ ds\ ,
\end{equation}
where $\Im$ indicates the imaginary part. Adding the contributions in (\ref{kin1}) and  (\ref{kin2}) the new expression for the kinetic
part of the Wigner equation is obtained
\begin{equation}\label{kintot}
\frac{\partial f_{wc}}{\partial t}\Bigg\vert_{kin}=-\frac{p}{m}\frac{\partial f_{wc}}{\partial x}+\frac{\hbar}{m}\frac{1}{\lambda}\frac{\partial}{\partial x}2\Im\int_0^{\infty}\ e^{-s/\lambda}e^{ips/\hbar}\ \psi(-)\psi^*(+)\ ds\ .
\end{equation}

As it regards the potential term of the Hamiltonian in (\ref{ham}), its elaboration is not influenced by the finite coherence length, and we report here
the final result only:~\footnote{In the derivation of Eq. (\ref{pot}) the damping factor $e^{-|s|/\lambda}$ has been attributed to the WF to obtain
an expression in terms of $f_{wc}$. With an equivalent procedure the damping factor could be attributed to the potential difference
$\left[V(+)-V(-)\right]$, thus leading to a damped potential ${\cal V}_{wc}$. In other words, it can be proved that: 
$\int {\cal V}_{w}\ f_{wc}\ dp'=\int {\cal V}_{wc}\ f_w\ dp'$. 
This means that the potential term in (\ref{pot}) does not contribute to $\partial f_{wc}/\partial t$ in points $x$ far from the region
where the potential $V(x)$ is not constant (e.g. for a potential step). A close analysis of the integral in the form given in Eq.(\ref{pot}) shows that
the integrand in such points has negligible contributions because of the very fast oscillations in ${\cal V}_w$ .} 
\begin{equation}\label{pot}
\frac{\partial f_{wc}}{\partial t}\Bigg\vert_{pot}=\frac{1}{\hbar}\int_{-\infty}^{\infty}{\cal V}_w(x,p-p')\ f_{wc}(x,p')\ dp'\ ,
\end{equation}
where
\begin{equation}
{\cal V}_w(x,p-p')=\frac{1}{2\pi i\hbar}\int_{-\infty}^{\infty} e^{-i(p-p')s/\hbar}\left[V(+)-V(-)\right]\ ds\ .
\end{equation}
Now using the definition of the Wigner equation together with (\ref{kintot}) and (\ref{pot}) we get:
\begin{equation}\label{eqfwc}
\frac{\partial f_{wc}}{\partial t}+\frac{p}{m}\frac{\partial f_{wc}}{\partial x}-\frac{1}{\hbar}\int_{-\infty}^{\infty}{\cal V}_w(x,p-p')\ f_{wc}(x,p')\ dp'
=g(x,p)\ ,
\end{equation}
where
\begin{equation}\label{corr}
g(x,p)=\frac{p_0}{m}\frac{\partial}{\partial x}2\Im\int_0^{\infty}e^{i(p+ip_0)s/\hbar}\ \psi(-)\psi^*(+)\ ds
\end{equation}
is the corrective term to the standard Wigner equation due to the finite coherence length. Here we have set $p_0=\hbar/\lambda$.
Note that Eqs. (\ref{eqfwc}) and (\ref{corr}) do not lead to a closed equation for $f_{wc}$ since they still contain the wavefunctions. 
On the other hand, as it will be shown in the following, a closed equation can be derived 
for a new function obtained as a generalization of $f_{wc}$.

\subsection{Dynamical equation}

In order to get a more compact analytical theory, we found it convenient to generalize the definition for the Wigner function $f_{wc}$. 
First we rearrange the definition of $f_{wc}$ in (\ref{WFc}) as
\begin{equation}\label{WFc2}
f_{wc}(x,p)=2\Re\int_{0}^{\infty}\ e^{i(p+ip_0)s/\hbar}\ \psi(-)\psi^*(+)\ ds\ ,
\end{equation}
where  $\Re$ indicates the real part. It should be noticed that from this expression it results that the introduction of the finite coherence length is represented mathematically by the shift of the Wigner momentum off the real axis.

Let us now define the new function 
\begin{equation}\label{ftilde}
\tilde{f}(x,\tilde{p})\equiv 2 \int_{0}^{\infty}\ e^{i\tilde{p}s/\hbar}\ \psi(-)\psi^*(+)\ ds=\tilde{f}_r+i\tilde{f}_i\ ,
\end{equation}
where $\tilde{p}\equiv p+ip_0$ and $\tilde{f}_r$ and $i\tilde{f}_i$  are the real and imaginary parts of $\tilde{f}$. Note that $f_{wc}=\tilde{f}_r$.
With the above definition, equation (\ref{eqfwc}) can be rewritten as
\begin{equation}\label{neq0}
\frac{\partial\tilde{f}_r}{\partial t}+\frac{p}{m}\frac{\partial\tilde{f}_r}{\partial x}-\frac{p_0}{m}\frac{\partial\tilde{f}_i}{\partial x}
-\frac{1}{\hbar}\int_{-\infty}^{\infty}{\cal V}_w(x,p-p')\ \tilde{f}_r(x,p')\ dp'=0\ .
\end{equation}
The above integral must be performed on the complex variable $\tilde{p}$ along the trajectory parallel to the real axis placed at a 
distance $p_0$ from the axis itself. Only the real part of $\tilde{p}$ varies along the integration path, as indicated in the
integral expression.\\ \noindent
Starting from the definition (\ref{ftilde}) and following again the standard procedure, it is possible to derive a closed dynamical equation for $\tilde{f}$. 
In particular the kinetic contribution after integration by parts reads:

\begin{eqnarray}
\hspace{-30 pt}\frac{\partial\tilde{f}}{\partial t}\Bigg\vert_{kin}&=&4i\frac{\hbar}{m}\ e^{i\tilde{p}s/\hbar}\ 
\left[\left(\frac{\partial}{\partial s}\psi(-)\right)\psi^*(+)-\psi(-)\left(\frac{\partial}{\partial s}\psi^*(+)\right)\right]\Bigg\vert_0^{\infty}\nonumber\\ 
&-&4i\frac{\hbar}{m}\int_0^{\infty}\ \frac{i\tilde{p}}{\hbar}\ e^{i\tilde{p}s/\hbar}\ 
\left[\left(\frac{\partial}{\partial s}\psi(-)\right)\psi^*(+)-\psi(-)\left(\frac{\partial}{\partial s}\psi^*(+)\right)\right]\ ds\nonumber\\ 
&=&2i\frac{\hbar}{m}\frac{\partial}{\partial x}\left(\psi(x)\psi^*(x)\right)-2\frac{\tilde{p}}{m}\frac{\partial}{\partial x}
\int_0^{\infty}\ e^{i\tilde{p}s/\hbar}\ \psi(-)\psi^*(+)\ ds\ .
\end{eqnarray}
Finally we get:
\begin{equation}
\frac{\partial\tilde{f}}{\partial t}\Bigg\vert_{kin}=2i\frac{\hbar}{m}\frac{\partial}{\partial x}\left\vert\psi(x)\right\vert^2-
\frac{\tilde p}{m}\frac{\partial\tilde f}{\partial x}\ .
\end{equation}
The evaluation of the potential contribution is straightforward, and the equation for  $\tilde f$ results to be
\begin{equation}\label{ftildeeq}
\frac{\partial\tilde{f}}{\partial t}+\frac{\tilde p}{m}\frac{\partial\tilde f}{\partial x}-2i\frac{\hbar}{m}\frac{\partial}{\partial x}\left\vert\psi(x)\right\vert^2=
\frac{1}{\hbar}\int_{-\infty}^{\infty}{\cal V}_w(x,p-p')\ \tilde{f}(x,p')\ dp'\ .
\end{equation}
Note that the real part of (\ref{ftildeeq}) returns (\ref{neq0}). 

To write equation (\ref{ftildeeq}) in a closed form we show that the following relation holds:
\begin{equation}
\int_\infty^\infty \tilde{f}\ dp=\vert\psi(x)\vert^2\ .
\end{equation}
In fact, from the definition of $f_{wc}=\tilde{f}_r$ in (\ref{WFc}), the integral of $\tilde{f}_r$ over the momentum variable
yields $\vert\psi(x)\vert^2$ (as discussed also in the next section). Now, using the definition 
$\tilde{f}_i=1/(2i)(\tilde{f}-\tilde{f}^*)$, straightforward calculations show that $\int \tilde{f}_i\ dp=0$, thus proving the above 
equality. Eq. (\ref{ftildeeq}) can finally be written in a closed form as:
\begin{equation}\label{neweq}
\hspace{-40 pt}\frac{\partial\tilde{f}}{\partial t}+\frac{\tilde p}{m}\frac{\partial\tilde f}{\partial x}-2i\frac{\hbar}{m}\frac{\partial}{\partial x}\int \tilde{f}(x,p')\ dp'=
\frac{1}{\hbar}\int_{-\infty}^{\infty}{\cal V}_w(x,p-p')\ \tilde{f}(x,p')\ dp'\ .
\end{equation}

\section{Discussion}
\subsection{Mean values}
The definition of the WF with finite coherence length given in (\ref{WFc}) is such that some of the known properties of the WF must be reconsidered.
The introduction of the exponential damping factor in the real-space variable preserves  the mean values of quantities which are local
in space. In particular it is still true that
\begin{equation}\label{psix2}
n_{\lambda}(x)\equiv\frac{1}{h}\int f_{wc}(x,p) dp=\vert\psi(x)\vert^2=n(x)\ ,
\end{equation}
as can be immediately seen from the definition in (\ref{WFc}) since the integration over $p$ yields the $\delta$ function in $s=0$.
Thus the integration over $p$ of the damped $f_{wc}$ leads to the standard particle density since $n(x)$ is a local quantity
and is not affected by a change in the space correlations.
On the contrary, since the particle momentum is related to the space correlations,
the integration over $x$ does not yield the standard result $\vert\phi(p)\vert^2$, but leads to a new expression
with an interesting physical interpretation. Following the usual procedure in the integral of $f_{wc}$ over $x$, we introduce the
$\phi(p)$'s as Fourier transforms of the $\psi(x)$'s:
\begin{eqnarray}
&&\hspace{-60 pt}n_{\lambda}(p)\equiv\frac{1}{h}\int f_{wc}(x,p)\ dx=
\frac{1}{h}\frac{1}{2\pi\hbar}\int\int\int\int e^{-ips/\hbar}e^{-p_0\vert s\vert/\hbar} \nonumber \\
&&\hspace{30 pt}\times e^{ip'(x+s/2)/\hbar}\ \phi(p')\ 
 e^{-ip''(x-s/2)/\hbar}\ \phi^*(p'')\ dp'\ dp''\ ds\ dx  \nonumber \\
&&\hspace{-60 pt}=\frac{1}{h}\int\int\phi(p')\ \phi^*(p'')\ dp'\ dp''\ 
\int e^{-i(p-\frac{p'+p''}{2})s/\hbar}\ e^{-p_0\vert s\vert/\hbar}\ ds \left(\frac{1}{2\pi\hbar}\int  e^{i(p'-p'')x/\hbar}\ dx\right) \nonumber \\
&&\hspace{-60 pt}=\frac{1}{h}\int\vert\phi(p')\vert^2\ dp'\ \left(\int_{-\infty}^0 e^{-i(p-p')s/\hbar} e^{p_0 s/\hbar}\ ds \right. 
\left.+\int_0^{\infty} e^{-i(p-p')s/\hbar} e^{-p_0 s/\hbar}\ ds\right)\ ,
\end{eqnarray}
where the integral definition of the $\delta(p'-p'')$ has been used, and the integration over $s$ has been split to treat the $\vert s\vert$ 
dependence. Now performing the integrations over $s$ leads to:
\begin{equation}\label{phip2}
n_{\lambda}(p)=\int\vert\phi(p')\vert^2\ \frac{1}{\pi}\ \frac{p_0}{p_0^2+(p-p')^2}\ dp'\ .
\end{equation}
The above expression represents a broadened particle density in momentum-space, where the broadening is a consequence of
assuming a finite value for the coherence length. In other words, for each $p$ the particle 
density in the momentum-space is weighted by a Lorentzian centred in $p$ of width $p_0=\hbar/\lambda$. 
When the coherence length goes to infinity the Lorentzian yields the $\delta(p-p')$, and the standard result is recovered. Expression
(\ref{phip2}) suggests the conclusion that while $\vert\phi(p)\vert^2$ is the momentum density in $p$ for the wavefunction with infinite 
coherence length, the integral in the l.h.s. of (\ref{phip2}) yields the momentum density $\langle n(p)\rangle_{\lambda}$ when the
coherence length is reduced to $\lambda$. 

Similar considerations hold when we consider the mean value of a physical quantity represented by the hermitian operator ${\cal A}$. 
With the standard definition of the WF we have:
\begin{equation}\label{Omv}
\langle{\cal A}\rangle=\frac{1}{h}\int\int f_{w}(x,p)\ A_w(x,p)\ dx\ dp\ ,
\end{equation}
where 
\[
A_w(x,p)=\int e^{-ips/\hbar}\ A(x+s/2,x-s/2)\ ds
\]
is the Weyl-Wigner transform of the operator ${\cal A}$. 

Let us first note that if ${\cal A}$ is a function $F({\sf x})$ of only the operator ${\sf x}$, then straightforward calculations show that
\begin{equation}\label{Fwx}
F_w(x,p)=F(x)\ .
\end{equation}
Similarly, if ${\cal A}$ is a function $G({\sf p})$ of only the operator $p$ then
\begin{equation}\label{Gwp}
G_w(x,p)=G(p)\ .
\end{equation}
Thus, combining (\ref{Fwx}) with (\ref{psix2}) we obtain:
\begin{equation}
\langle F({\sf x})\rangle=\frac{1}{h}\int\int f_w(x,p)F(x)\ dx\ dp=\int n(x) F(x) dx\ ,
\end{equation}
and, using (\ref{psix2}),
\begin{equation}
\langle F({\sf x})\rangle_{\lambda}\equiv\frac{1}{h}\int\int f_{wc}(x,p)F(x)\ dx\ dp=\int n_{\lambda}(x) F(x) dx=\langle F({\sf x})\rangle\ .
\end{equation}
Analogously, combining (\ref{Gwp}) with (\ref{phip2})
\begin{equation}
\langle G({\sf p})\rangle=\frac{1}{h}\int\int f_w(x,p) G(p)\ dx\ dp=\int n(p) G(p) dp\ ,
\end{equation}
and
\begin{equation}
\langle G({\sf p})\rangle_{\lambda}\equiv\frac{1}{h}\int\int f_{wc}(x,p) G(p)\ dx\ dp=\int n_{\lambda}(p) G(p) dp\ .
\end{equation}
The above results suggest to define the mean value of a general observable ${\cal A}$ in the state with finite coherence 
length as
\begin{equation}
\langle{\cal A}\rangle_{\lambda}\equiv\frac{1}{h}\int\int f_{wc}(x,p)A_w(x,p)dx\ dp\ .
\end{equation}
The mean value of the observable ${\cal A}$ in the coherent state $\psi(x)$ can still be obtained as follows:
\begin{equation}
\langle{\cal A}\rangle=\int\int \psi^*(x_1)\ A(x_1,x_2)\ \psi(x_2)\ dx_1\ dx_2\ ,
\end{equation}
and moving to the central $x=(x_1+x_2)/2$ and difference $s=x_2-x_1$ variables, we get:
\begin{equation}
\langle{\cal A}\rangle=\int\int \psi^*(x-s/2)\ A(x-s/2,x+s/2)\ \psi(x+s/2)\ ds\ dx\ .
\end{equation}
Inserting the $\delta(s-s')$ and integrating over $s'$ we can write:
\begin{eqnarray}
\langle{\cal A}\rangle&=&\int\int\int e^{-p_0\vert s\vert/\hbar}\psi^*(x-s/2)\ \psi(x+s/2) \nonumber \\
&&\times\ e^{p_0\vert s'\vert/\hbar}\ A(x-s'/2,x+s'/2)\ \delta(s-s')\ \ ds\ ds'\ dx\ ,
\end{eqnarray}
and then using the plane-wave representation for the $\delta$ function, we obtain
\begin{eqnarray}
\langle{\cal A}\rangle&=\frac{1}{2\pi\hbar}\int\int\int\int e^{-ips/\hbar}\ e^{-p_0\vert s\vert/\hbar}\psi^*(x-s/2)\ \psi(x+s/2) \nonumber \\
&\times\ e^{ips'/\hbar}\ e^{p_0\vert s'\vert/\hbar}\ A(x-s'/2,x+s'/2)\ ds\ ds'\ dx\ dp\ .
\end{eqnarray}
Recalling (\ref{WFc}), we finally get:
\begin{equation}\label{Omvc}
\hspace{-40 pt}\langle{\cal A}\rangle=\frac{1}{h}\int\int\int\ f_{wc}(x,p)\ e^{ips'/\hbar}\ e^{p_0\vert s'\vert/\hbar}\ A(x-s'/2,x+s'/2)\ ds'\ dx\ dp\ .
\end{equation}
The formal difference between (\ref{Omv}) and (\ref{Omvc}) is that for the latter we cannot define an analogous of
$A_w(x,p)$, since in (\ref{Omvc}) the integral over $s'$, taken by itself, is divergent. On the other hand the integration of $f_{wc}$ 
over $x$ balance the divergence, and the expression, as a whole, is well defined and finite.

\begin{figure}[htpb]
  \begin{center}
 \includegraphics*[width=0.8\linewidth,angle=270]{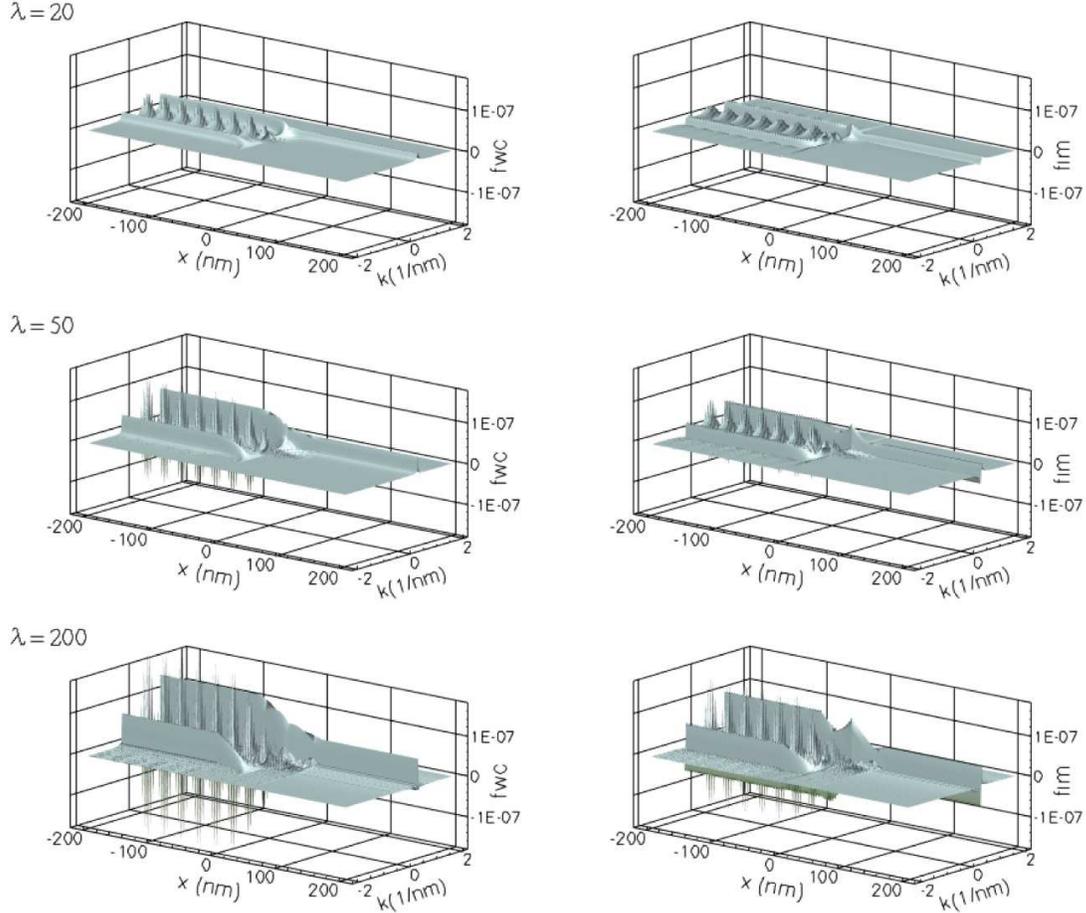}
    \caption{\label{fig1} (Color online) Real (left column) and imaginary part (right column) of the Wigner function $\tilde f$ with finite 
    coherence length for the case of a potential barrier 0.4 eV high, and 2 nm wide. The energy of the scattering state is 0.3 eV.
  The different lines correspond to the indicated coherence lengths.}
 \end{center}
 \end{figure}

\subsection{Limit of the dynamical equation for infinite coherence length}
An important point that can now be raised is wether the limit of the new equation, as given in (\ref{neweq}), returns the standard Wigner 
equation when $\lambda\to\infty$, i.e. when $p_0\to 0$. From (\ref{neq0}) we see that this is actually the case if
\begin{equation}\label{limit}
\lim_{p_0\to 0} p_0\frac{\partial\tilde{f}_i}{\partial x}=0\ .
\end{equation}
The above condition is satisfied as long as $\partial\tilde{f}_i/\partial x$ remains finite or diverges slower then $1/p_0$ when $p_0\to 0$.
In the case of the finite potential barrier discussed in the following, we have certified numerically that the condition (\ref{limit}) is verified.
Besides, Rossi and coworkers \cite{rosati13} have shown that for a $\delta$-like potential barrier the standard Wigner equation is verified
by the WF obtained with the analytical expression of the scattering states.

\subsection{Numerical examples}
As a numerical example we consider the typical case of a rectangular potential barrier of hight $V_0$ and thickness $\ell$. A scattering
state
\begin{equation}\label{ss}
\Psi(x)=\left\{ \begin{array}{ll}
Ae^{ikx}+Be^{-ikx} & x<-\ell/2 \\
Ce^{\kappa x}+De^{-\kappa x} & -\ell/2<x<\ell/2 \\
Ee^{ikx} & x>\ell/2
\end{array} \ .
\right.
\end{equation}
is considered, with energy $\varepsilon<V_0$, where:
\[
k=\sqrt{\frac{2m\varepsilon}{\hbar^2}}\ \ \ ,\ \ \ \kappa=\sqrt{\frac{2m(V_0-\varepsilon)}{\hbar^2}}\ .
\]
The WF can be calculated analytically for both the infinite and finite coherence lengths, even though the results are rather
cumbersome (see for example \cite{rosati13}). Fig.~(1) shows the function $f_{wc}$ as real part of the $\tilde{f}$ as well as the imaginary
part of the same function for different values of $\lambda$. For increasing coherence length $f_{wc}$ approaches the combination of 
$\delta$-like functions obtained by the coherent scattering states, plus the contribution of the correlations between the incoming and
the reflected waves.

\section{Conclusions}
The theory of the WF has been revised to include the effect of a finite coherence length, corresponding to move the Wigner momentum
$p$ off the real axis. 

A new function $f_{wc}$ is defined as in (\ref{WFc}), and the main results obtained with such revision are expressed by the 
new dynamic equations in (\ref{neweq}).

In terms of the new WF, $f_{wc}$, mean values of physical quantities can be calculated. In particular mean values of the quantities
depending only on the position $x$ are unaltered, while mean values of physical quantities depending on the momentum 
must account for a broadening of the momentum distribution according to a Lorentian distribution, as in (\ref{phip2}).

The limit of the new equation as the coherence length goes to infinity is also analyzed, obtaining the conditions which allow to 
recover the standard Wigner equation. 
For the sake of concreteness, a numerical example has been developed using the standard case of study of a rectangular potential
barrier.

The theory developed in this work can be applied to study the dynamics of the WF, in absence of scatterings, avoiding divergence 
problems by using a suitably large coherence length..

\ack
The authors are extremely grateful to Fausto Rossi,  Lucio Demeio and Alberto Guandalini for stimulating and fruitful discussions.

\section*{References}
%\bibliography{References}
\providecommand{\newblock}{}

%\begin{thebibliography}{10}
%\bibitem{book1} Goosens M, Rahtz S and Mittelbach F 1997 {\it The \LaTeX\ Graphics Companion\/} 
%(Reading, MA: Addison-Wesley)
%\bibitem{eps} Reckdahl K 1997 {\it Using Imported Graphics in \LaTeX\ } (search CTAN for the file `epslatex.pdf')
%\end{thebibliography}

\end{document}